# Mechanical Properties of Norway Spruce: Intra-Ring Variation and Generic Behavior of Earlywood and Latewood until Failure

Christian Lanvermann,* Philipp Hass, Falk K. Wittel, and Peter Niemz

The alternating earlywood and latewood growth ring structure has a strong influence on the mechanical performance of Norway spruce. In the current study, tensile tests in the longitudinal and tangential directions were performed on a series of specimens representing one growth ring at varying relative humidities. All tested mechanical parameters, namely modulus of elasticity and ultimate tensile stress, followed the density distribution in the growth ring, with the minimum values in earlywood and the maximum values in latewood. The samples were conditioned at three the relative humidities 50%, 65% and 95%. With increasing relative humidity, the values of the mechanical parameters were found to decrease. However, due to the high local variability, this decrease was not statistically significant. The test in the tangential direction on a set of earlywood and latewood specimens at 65% relative humidity revealed a similar limit of linear elasticity for both early- and latewood. Where the strength of both tissues was equal, the strain at failure was significantly greater for earlywood. Furthermore, the portion of the non-linear stress/strain behavior for earlywood was significantly greater. A Weibull analysis on the ultimate tensile strength revealed a tissue-independent Weibull modulus, which indicates similar defect distributions. For both, the failure occurred in the middle lamella.

*Contact information: ETH Zurich, Institute for Building Materials, Schafmattstrasse 6, 8093 Zurich, Switzerland, *Corresponding author: lanvermannchr@ethz.ch*

## INTRODUCTION

Wood is a cellular and anisotropic material with a strict hierarchical arrangement of its constituents. The anisotropic mechanical behavior with respect to the principal directions (longitudinal (L), radial (R) and tangential (T)) of bulk wood is well known (Keunecke *et al.* 2008; Neuhaus 1983;; Wagenführ 2000). On the growth ring scale, the structure of softwoods with their alternating high-density latewood (LW) and low-density earlywood (EW) bands (*e.g.*, Lanvermann *et al.* 2013b) adds a further level of complexity. The different tissues exhibit different transverse physical properties with respect to moisture-induced deformation (Derome *et al.* 2011; Derome *et al.* 2013 Keunecke *et al.* 2012; Rafsanjani *et al.* 2012) and mechanical properties as derived from simulations (Kahle and Woodhouse 1994; Modén and Berglund 2008b; Persson 2000; Rafsanjani *et al.* 2012). Due to its hierarchical structure, it is evident that the macroscopic behavior originates from features on a microscopic scale. For example, when considering wood drying, defects on the cell wall level as monitored by acoustic emission can lead to macroscopic defects if the drying process is not properly controlled (Rosner 2012). Furthermore, collapse of wooden structures can be initiated by transverse failure at cell level (Gustafsson 2003). Therefore, knowledge of local mechanical parameters and the use of advanced modeling techniques can shed light on the underlying mechanisms. The knowledge of these mechanisms can then lead to appropriate process control and a better utilization of the wood material, as well as new insights for improved hierarchical multi-scale models for wood.

Experimental studies on local mechanical parameters involve investigations on single fibers (Eder *et al.* 2009) and separated EW and LW sections (Moon *et al.* 2010) as well as observation of the different behaviors of intact tissue (Farruggia and Perré 2000; Jernkvist and Thuvander 2001; Modén and Berglund 2008a; Sinn *et al.* 2001), but neglect the well-known influence of moisture content (MC) on the mechanical properties (Gerhards 1982; Neuhaus 1983; Ekevad and Axelsson 2012; Naylor *et al.* 2012).

Failure and crack propagation observations in the RT plane (*e.g.*, Dill-Langer *et al.* 2002; Fruehmann *et al.* 2003; Hass *et al.* 2012; Wittel *et al.* 2005) reveal the failure mechanisms associated with EW and LW. What is observed in LW is a brittle failure with a fracture along the cell wall interface; in EW, cell wall rupture in the thin-walled cells can be seen. Because the parameters have such a large variability, as observed even within the individual tissues in a growth ring (Modén and Berglund 2008a), a statistical approach is needed to describe them. One such approach is the extreme value distribution function proposed by Weibull (1951), which is based on the weakest-link theory and has been successfully applied to ceramics (Nanjangud *et al.* 1995), metals (Nyahumwa 2005), and wood beams (Danielsson and Gustafsson 2010; Aicher *et al.* 2007) to describe brittle material failure.

The aim of the current study is twofold: To study the mechanical parameters (modulus of elasticity (MOE) and ultimate tensile strength (UTS)) in the T direction and MOE in the L direction within growth rings of Norway spruce for different ambient relative humidities (RHs), and to elaborate the differences in the mechanical behavior (MOE, UTS, and yield point) at constant RH of EW and LW in T until failure.

**EXPERIMENTAL**

**Materials and Methods**

The material for the current investigations originated from mature wood approximately 1 m above the ground, which is the same stem of Norway spruce that was previously characterized by Lanvermann *et al.* (2013b). In the current approach, the mechanical behavior of several series of samples, each representing one single growth ring, and the different behaviors of EW and LW was studied. Rectangular solid pieces of 9-mm width and 47-mm length were cut from a board, with their longest dimensions oriented parallel to and perpendicular to the L direction, respectively (see Fig. 1a).

The rectangular solids were first submerged in water for at least two weeks under mild vacuum conditions using a water-jet pump until they were completely saturated. Using a sliding microtome, a series of consecutive specimens was prepared, each representting one growth ring. Previous investigations of the cross-sectional cell geometry on growth rings of this particular stem showed that the ratio of the radial diameter of an EW and a LW cell is *ca.* 2:1 (Lanvermann *et al.* 2013b). In the T direction, cell dimensions are more or less constant as a direct consequence of wood formation, in which a radial row originates from one individual cambium cell (Lanvermann *et al.* 2013b).

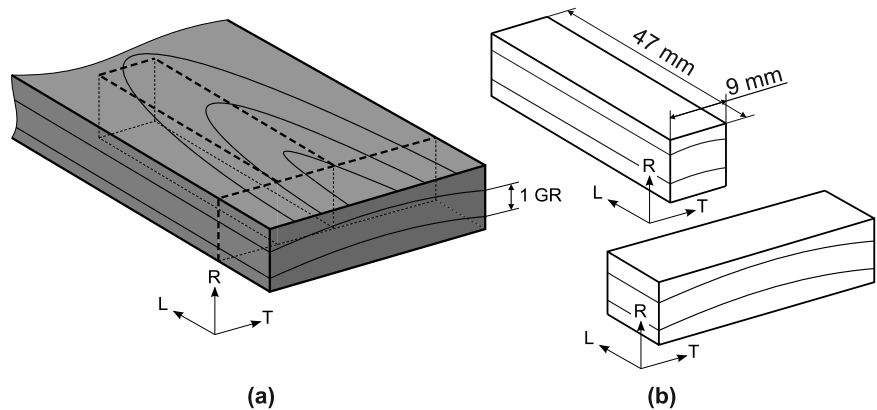

**Fig. 1.** Schematic of sample preparation. Orientation of the cuboids within the board (a) and local material orientation in the longitudinal and tangential sample configuration (b). The growth ring curvature in (a) is exaggerated.

The sample thickness was adapted to account for the smaller cell dimensions in the R direction of LW. Preliminary tensile tests with EW tested in the T direction showed that a thickness of 300 μm leads to forces of a sufficient order of magnitude to be measured. The abovementioned ratio was then applied, leading to an average sample thickness for EW of 282.9 μm (CoV: 0.115) and 179.7 μm (CoV: 0.286) for LW. Therefore, the number of cells and thus number of failure points in the stressed region was comparable for EW and LW. The classification of the samples as EW or LW was attained by visual inspection during cutting. A mixture of glycerin, alcohol, and water was applied to facilitate cutting with the microtome and afterwards removed by washing the samples with alcohol. The sample series were then stored for at least one week in climate-controlled rooms with nominal RHs of 50%, 65%, and 95%. To prevent warping of the samples, they were placed between microscope glass slides that were wrapped with adhesive tape. In total, 12 series of samples were prepared so that each growth ring was tested twice per RH level in the L as well as the T direction.

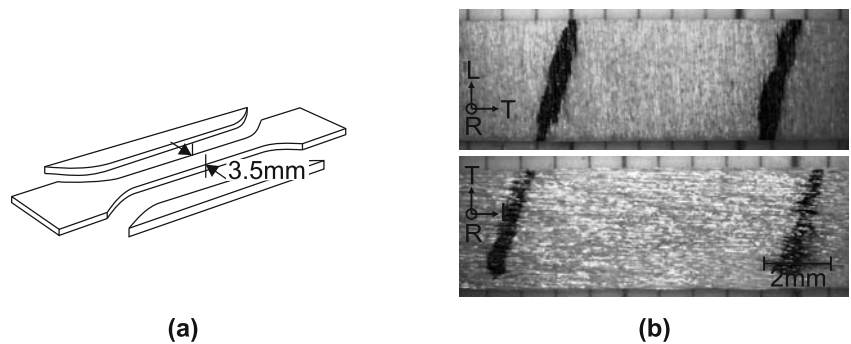

**Fig. 2.** Schematic of sample preparation. Rectangular solid pieces (a) cut into slices and punched into dog-bone shape. Tensile tests performed with the load applied in tangential direction ((b) upper image) and longitudinal direction ((b) lower image).

In addition to the sample series, a batch of EW and LW tangential samples of the aforementioned thicknesses was prepared and stored at 65% RH. Because previous investigations showed that the gravimetric MC is constant throughout the growth ring (Dvinskikh *et al.* 2011; Lanvermann *et al.* 2013a), additional rectangular solid pieces were equilibrated at the different RH levels and then dried at 103 °C to determine the MC. This approach avoided the problems that the drying step could cause (*i.e.*, warping and cracks). The tensile tests were performed on a micro-test stage (Deben, UK) equipped with a load cell with a maximum nominal capacity of 300 N at a testing speed of 0.5 mm min$^{-1}$. The samples were trimmed to their final dog-bone shape by means of a custom stamping template with razor blades directly before testing (see Fig. 2a). The displacement measurement during testing was accomplished without contact using an optical strain measurement system (Correlated Solutions, USA) while the images were recorded with a CCD camera (Allied Vision Technology, Germany) mounted on a stereo microscope (Olympus, Japan) to enable sufficient magnification so that the region for strain measurement filled nearly the whole image (about 109 pixel/mm). Two pen markings on the samples' surface served as contrast markings for deformation measurement (see Fig. 2b). During testing, the ambient conditions (temperature and RH) were recorded using two climate sensors (Almemo, Germany) adjacent to the testing machine.

The representative density distribution within the growth rings was determined gravimetrically at non-climate controlled conditions at 22.9 °C and 45.3% RH. The sample weights were recorded using a precision scale (Ohaus, USA, 0.0001g precision), and the sample thickness was measured with a thickness gauge (Mitutoyo, Japan, 0.001 mm precision). The planar sample area was determined optically by thresholding an orthographic image. The appropriate upper and lower thresholding limits were selected manually from the gray value histogram. According to these limits, the image was binarized and the number of pixels for the sample area was multiplied by the corresponding pixel size.

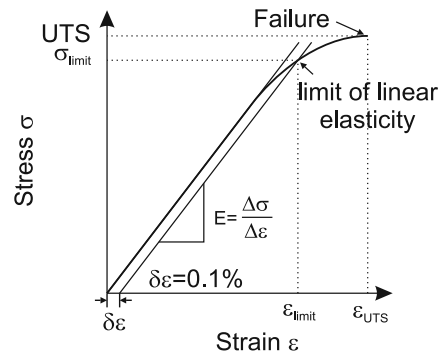

**Fig. 3.** Schematic stress-strain curve. MOE was determined within 20 to 40% of the failure stress ($\sigma_{UTS}$). The limit of linear elasticity and the corresponding stress and strain were determined by applying an offset to the MOE of 0.1% strain.

The separate datasets of force and strain were passed to MATLAB® for further evaluation. Because the recording frequencies of the two measurement systems (strain and force) were different, the datasets had to be synchronized and resampled. The stress was calculated by dividing the force by the respective initial sample's cross-sectional area. The MOE was calculated by a linear regression within 20 to 40% of the UTS (Fig. 3). Furthermore, the limit of elasticity (stress and strain) were determined at 0.1% strain offset from the linear part of the stress/strain curve.

The Weibull analysis of the EW and LW samples was performed following the procedure as described in DIN EN 61649:2008). In order to depict the corresponding failure mechanisms in tangential tension, images of the failure surfaces were taken using an electron microscope (FEI Quanta, USA at 600x magnification).

## RESULTS AND DISCUSSION

### Variation of Mechanical Properties within Growth rings

The intra-ring variation of the mechanical and physical properties, namely $MOE_L$, $MOE_T$, and $UTS_T$, where the subscripts refer to longitudinal and tangential directions, and density, for a RH level of 50%, are given in Fig. 4. The data represent the results of two sample series containing the same growth ring. An electron micrograph in the lower part of the figure illustrates the varied sample thicknesses for EW and LW (Fig. 4). The radial growth ring position was normalized, where 0 denotes EW and 1 is LW. Furthermore, analytical functions were fitted to the data, whose parameters are given in Table 1. As clearly seen, all parameters followed the same trend, with the lowest values in EW and the highest values in LW. The EW and LW densities of around 330 and 741 kg m$^{-3}$ lay within the range reported in a previous study on the same stem (278 to 344 kg m$^{-3}$ (EW) and 596 to 727 kg m$^{-3}$ (LW) and a mean density of 353 kg m$^{-3}$) (Lanvermann *et al.* 2013b). $MOE_L$, with values of around 7500 MPa for EW and 20600 MPa for LW, were higher than those reported in the literature for single fibers based on cell wall area (3000 MPa for EW and 15000 MPa for LW (Eder *et al.* 2009)) and lower than those predicted from a hierarchical model for intact tissue (33200 MPa for EW and 43000 MPa for LW (Persson 2000)). However, it has to be clearly stated that the data of Persson were modeled based on a mean density of 400 kg m$^{-3}$, which clearly deviates from the mean wood density as used in this experiment and thus can partly explain the higher values of Persson. A similar behavior could be found for $MOE_T$. While in the present study $MOE_T$ was around 106 MPa for EW and 950 MPa for LW, studies on intact tissue using digital speckle photography reached values of 65 to 400 MPa for EW and around 1200 to 4000 MPa for LW (Farruggia and Perré 2000; Jernkvist and Thuvander 2001; Modén and Berglund 2008b). For $UTS_T$, there are no values available in the literature that can be readily compared. Nevertheless, the measured $UTS_T$ data (2.7 MPa for EW and 3.8 MPa for LW) were in accordance with the range reported for perpendicular-to-grain tensile loading of bulk wood (1.5 to 4.0 MPa (*e.g.* Wagenführ 2000)). Generally, a comparison with literature data can only provide an order of magnitude estimate, considering the different sample geometries used (individual fibers *vs*. tissue slices *vs*. intact tissue *vs*. bulk wood) which involve different degrees of interplay of the individual cells that affect the determined values.

**Table 1.** Parameters of the Fit Functions for the Individual Properties at 9.28% MC

| Direction | Property | | $R^2$ | Parameter | | | |
|---|---|---|---|---|---|---|---|
| | | | | a | b | c | d |
| L | MOE | (MPa) | 0.864 | 4.415 x 10$^3$ | 1.024 | 2.676 | 8.636 |
| T | MOE | (MPa) | 0.888 | 6.931 x 10$^1$ | 1.239 | 7.503 x 10$^{-13}$ | 3.708 x 10$^1$ |
| T | UTS | (MPa) | 0.804 | 4.068 x 10$^{-15}$ | 3.639 x 10$^1$ | 2.731 | -1.966 x 10$^{-1}$ |
| | Density | (kg m$^{-3}$) | 0.928 | 2.837 x 10$^2$ | 3.127 x 10$^{-1}$ | 5.992 x 10$^{-3}$ | 1.213 x 10$^1$ |

A function of the form $y = a * exp(b * x) + c * exp(d * x)$ was used.

In addition to the material density, which clearly has a major influence on the mechanical properties as shown in Fig. 4, differences in the mechanical properties between EW and LW are associated with different microfibril angles (MFA) of the secondary cell wall.

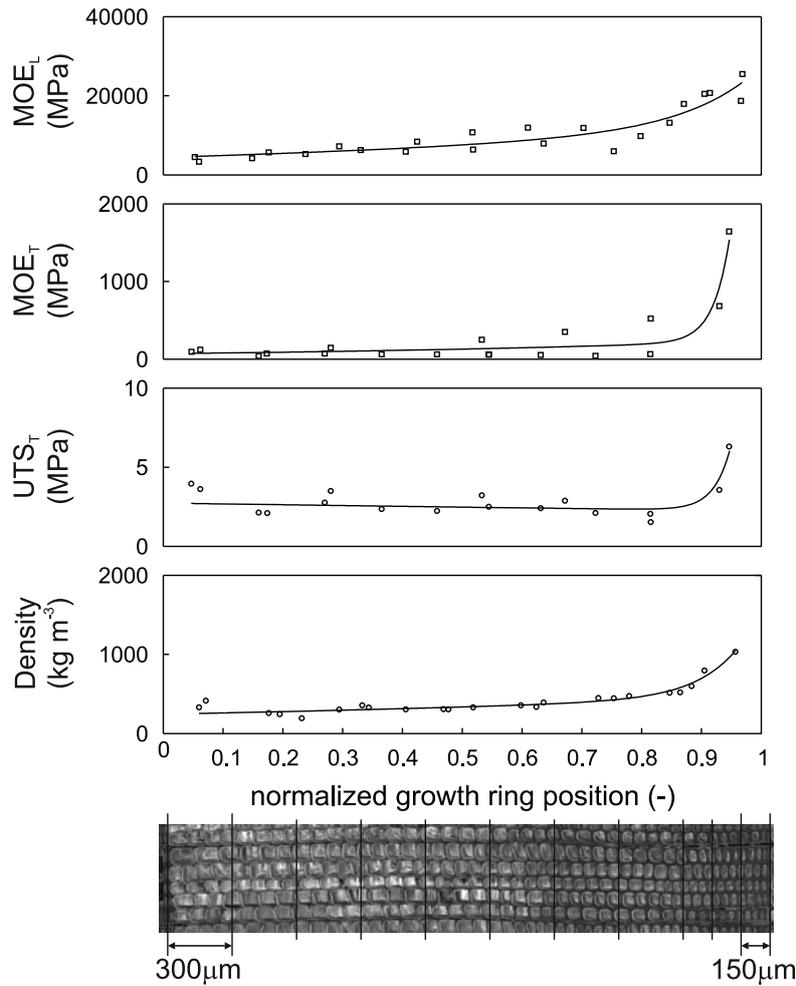

**Fig. 4.** Longitudinal and tangential MOE (MOE$_L$ and MOE$_T$), tangential UTS (UTS$_T$), and density of consecutive samples at 9.28% MC. The normalized growth ring position 0 denotes EW and 1 denotes LW. An electron micrograph illustrates the variation in sample thickness in 300 μm of EW samples and 150 μm of LW samples.

The MFA describes the inclination of the helically wound cellulose fibrils towards the longitudinal cell axis, and gradually decreases from EW (around 20°) to LW (around 12°) (*e.g.* Lanvermann et al. 2013b, Roszyk *et. al* 2013, Donaldson 2008). The influence of MFA on the longitudinal compliance is described in numerous investigations for compression wood with its altered chemical composition (*e.g.* Tarmian and Azadfallahm 2009), and its relatively high MFA (*e.g.* Evans and Ilic 2001; Burgert *et al.* 2004). However, the transverse behavior of the microfibrils is regarded as isotropic (*e.g.* Boutelje 1962), and therefore the influence of MFA on the transverse mechanical properties is rather limited (*e.g.* Astley *et al.* 1998).

The equilibrium moisture contents (EMCs) as determined for the individual RH values of 50, 65, and 95% RH were 9.3, 13.8 and 23.7%, respectively, and therefore were comparable to those reported in the literature, considering that the samples were conditioned in desorption from a fully saturated state (*e.g.* Skaar 1988). Following the definition of Persson (2000), the EW makes up approximately 63% of the whole growth ring, transitionwood (TW) makes up 30%, and LW makes up 7%.

Because the classification was accomplished visually in the current investigation, it was limited to EW and LW; TW was neglected and included in EW. With the high share of the growth ring, EW and TW make up 93%, there are far more EW specimens per growth ring than LW specimens (approximately 8 for EW and TW and 1 to 2 for LW). In assessing the influence of MC, only EW could be used in the present study because there were so few LW specimens per MC level.

The data for the two loading directions and mechanical properties are given in Figs. 5 and 6. The reduction of the MOE with increasing RH is well known (Gerhards 1982; Neuhaus 1983). According to these authors, a smaller impact in the L direction than in the T direction is observed. In the L direction, with the 9.3% MC level set as the reference state, measurements on bulk wood show a 15.6% reduction in MOE at 13.8% MC and a 20.7% reduction at 23.7% MC (calculated from data of Neuhaus (1983) for bulk wood).

In addition to the impact of MC, the different investigated sample geometries influence on the determined values. Where values for bulk wood represent the behavior that includes the interplay of both tissues, investigations on thin slices more or less represent the behavior of the isolated tissue without the influence of the other, provided that the testing was conducted at equilibrium MC.

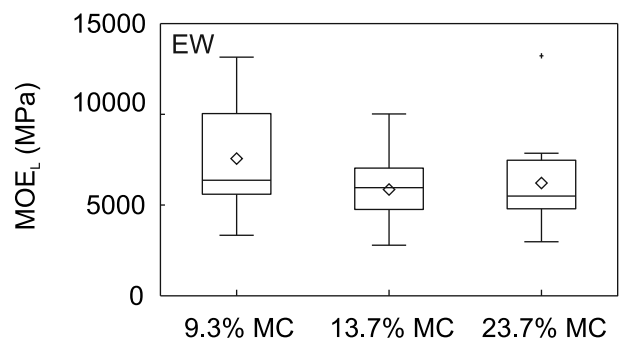

**Fig. 5.** Longitudinal MOE for EW and the three MC levels

When only the mean values of $MOE_L$ were considered, this ratio was approximately reflected in the current investigations (see Fig. 5 diamonds). However, when the whole dataset was taken into account, the considerable variability within the MC levels led to statistically insignificant differences between them. A similar behavior could be found for the MOE reduction in T, where a reduction of 16.4% (13.8% MC) and of 43.0% (23.7% MC) was found for bulk wood (Neuhaus 1983). The average $MOE_T$ in the current investigation clearly reflected this reduction. However, the reduction was significantly higher, 41.0% and 61.2%, respectively. Furthermore, it is worth noting that the variability decreased with increasing MC. A possible explanation of this decreased variability might be the reduction of eigenstresses due to the softening of the cell wall material with increasing MC and therefore a more homogeneous behavior. The tangential UTS shown in Fig. 5b showed a reduction of the mean of about 20% for the 23.7% MC level, while the 13.8% MC was equal to the 9.3% MC level, which was again not statistically different at the 5% confidence level.

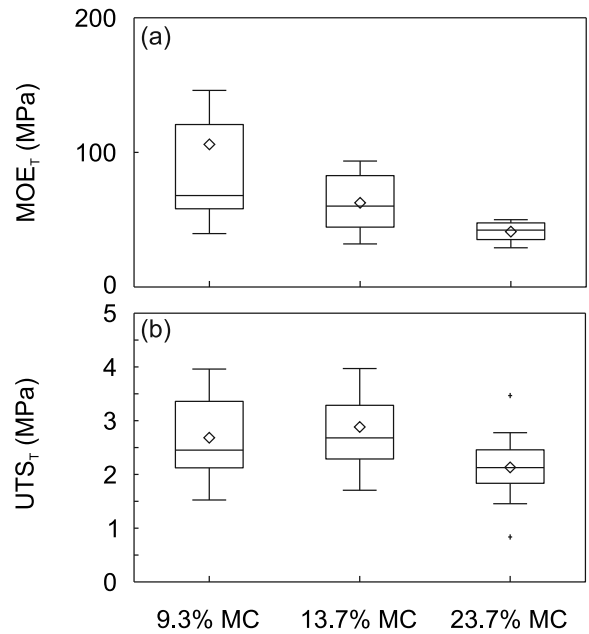

**Fig. 6.** Tangential MOE (a) and UTS (b) for EW and the three MC levels

## Generic Mechanical Behavior of EW and LW

The remarkable variability of the mechanical properties discussed in the previous section requires a statistical approach with more samples for both EW and LW to elaborate the mechanical behavior. Therefore, the concept of testing the same growth ring was dropped and a number of samples containing either EW or LW were prepared and tested in the T direction at the 65% RH level. The representative stress/strain curves, as given in Fig. 7, show a clear difference in the stress/strain behavior of the two tissues. Considering the mean values and the relatively high coefficients of variation (Table 2), the need of the application of a statistical approach becomes evident. The differences between EW and LW were analyzed using a two-way t-test. From the results given in Table 2, it was apparent that, whereas all strain-related quantities ($\varepsilon_{limit\ 0.1\ T}$, $\varepsilon_{UTS\ T}$ and $MOE_T$) were significantly different for EW and LW, the differences for all stress-related quantities ($\sigma_{limit\ 0.1\ T}$ and UTS) were not significant at a 95% confidence level.

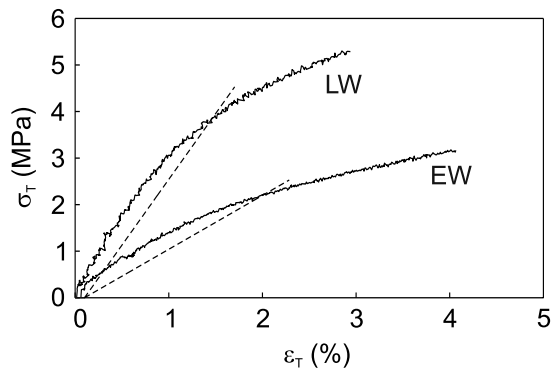

**Fig. 7.** Representative stress/strain curves until failure in tangential tension for EW and LW and corresponding MOE with an 0.1% strain-offset (dashed lines) to determine limit of linear elasticity

**Table 2.** Parameters Mean in Tangential Direction, Coefficient Of Variation (CoV) Given in Round parentheses

|  |  | EW |  | LW |  | Significance |
|---|---|---|---|---|---|---|
| N |  | 38 |  | 37 |  |  |
| $\varepsilon_{T\ limit\ 0.1}$ | (-) | 0.016 | (0.19) | 0.013 | (0.40) | ** (p=0.0094) |
| $\sigma_{T\ limit\ 0.1}$ | (MPa) | 2.061 | (0.40) | 2.461 | (0.56) | n.s. |
| $\varepsilon_{UTS\ T}$ | (-) | 0.046 | (0.37) | 0.027 | (0.67) | *** (p=8.92 $10^{-6}$) |
| $UTS_T$ | (MPa) | 4.39 | (0.48) | 4.73 | (0.46) | n.s. |
| $MOE_T$ | (MPa) | 146 | (0.55) | 334 | (0.71) | *** (p=1.40 $10^{-5}$) |

n.s. $p>0.05$; *: $p<0.05$; **: $p<0.01$; ***: $p<0.001$

The authors are aware that the offset for determining the limit of linear elasticity is more or less arbitrary (see Fig. 7). However, because an equal offset was applied throughout the evaluation, it served as a measure for the non-linear portion of the stress/strain relationship. Whereas the $\varepsilon_{limit\ 0.1\ T}$ of EW was slightly higher than that of LW, the $\sigma_{limit\ 0.1\ T}$ was slightly higher for LW, but the difference was not significant at a 95% confidence level because of the high variation of both parameters. While $\varepsilon_{UTS\ T}$ was significantly higher for EW, $UTS_T$ did not differ significantly for EW and LW. Furthermore, $MOE_T$ was significantly higher for LW, although it did not reach values as high as those reported in the literature (*e.g.* Farruggia and Perré 2000). The relatively low $MOE_T$ in LW was most likely caused by an inaccuracy in the visual classification into EW and LW that led to no pure LW being tested, but only a combination of both. It should be noted that all parameters were associated with relatively high coefficients of variation (Tab. 2).

The driving factor for the different stress-strain behavior of EW and LW in tangential tension is the geometry of the individual cells of the material (*e.g.* Rafsanjani *et al.* 2013; Kifetew 1999; Watanabe 1998). As illustrated by the micrograph in Fig. 4, the structure of EW can be considered a honeycomb structure, and LW is a brick-like structure. By loading a sample in the R direction in the linear elastic region and observing the strain field within a growth ring of Norway spruce, Modén and Berglund (2008a) identified the dominant deformation mechanism in the two cell types: cell wall bending in EW, and a combination of cell wall bending and cell wall stretching in LW. For EW, due to the more or less hexagonal shape of the cells it seems likely that the bending mechanism in the linear elastic region is transferable to the case when loading in T direction within the linear elastic region. For LW, the combination of cell wall bending and stretching is also likely when loading in T, although one could expect a higher share of cell wall stretching due to a higher share of cell wall material being oriented parallel to the loading direction. However, a significant structural difference when testing in T is the orientation of the ray tissue, which is loaded perpendicular to its axis. The role of rays in the overall deformation behavior for testing in T remains unknown. By means of sufficient imaging techniques, the cellular deformation within the linear elastic region could be observed in-situ in order to test the above hypotheses. In addition to the behavior in the linear elastic regime, it is evident from the stress/strain curves in Fig. 7 and the mean values in Table 2 that the non-linear stress/strain response of EW and LW is different. The corresponding deformation mechanisms, however, can only be subject of speculation with the current experimental approach.

Apart from the above mentioned limitations of the current experimental approach in identifying the deformation mechanisms of EW and LW, the tangential failure was investigated using a Weibull analysis. First, ln(UTS$_T$) was plotted against the failure probability and a regression was applied on a linear form of the Weibull distribution to visually test whether the UTS$_T$ data followed a Weibull distribution (Fig. 8a).

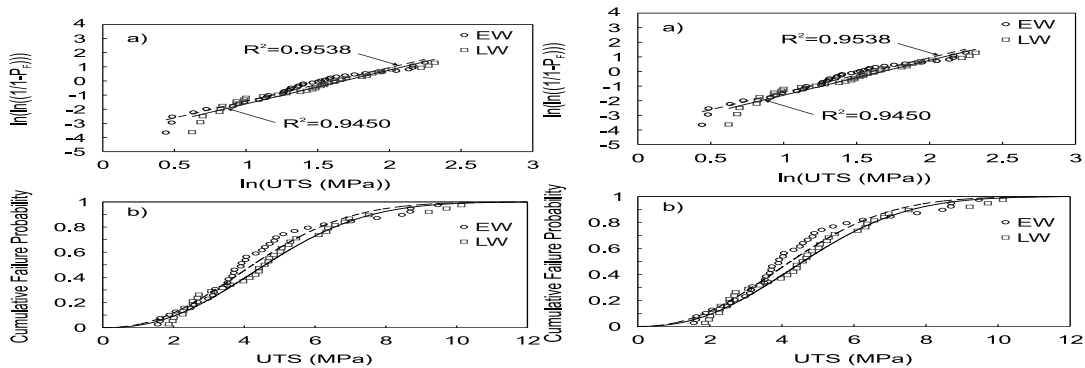

**Fig. 8. (a)** Weibull plot for EW (circles) and LW (grey rectangles) with linear regression assuming a single failure mechanism; **(b)** Cumulative failure probability for EW and LW using the two-parameter Weibull distribution with maximum-likelihood estimated parameters

As can be clearly seen, the data for EW and LW were close to the regression line (Fig. 8a). After this first visual test, an F-test was conducted, which did not reject the hypothesis that the data for EW and LW could be described by a Weibull distribution on a 95% confidence level. The cumulative failure probability for EW and LW are given in Fig. 8b and confirm the initial tests. The final maximum likelihood estimates and their 95% confidence bounds for the two-parameter Weibull distributions are given in Table 3.

**Table 3.** Weibull Analysis Parameters for EW and LW in Tangential tension.

|   | EW |  |  | LW |  |  |
|---|---|---|---|---|---|---|
|   |   | 95% confidence bound |  |   | 95% confidence bound |  |
|   |   | Lower | Upper |   | Lower | Upper |
| N | 38 | - | - | 37 | - | - |
| m | 4.9802 | 4.2960 | 5.7734 | 5.3587 | 4.6401 | 6.1886 |
| $\sigma_\theta$ | 2.2804 | 1.7982 | 2.8920 | 2.3696 | 1.8581 | 3.0218 |

It can be clearly seen that the Weibull modulus *m* of LW was slightly higher than that of EW. However, when the 95% confidence bounds of both parameters were considered, they were basically equal. The characteristic strength $\sigma_\theta$ for LW was also slightly higher than for EW, but when the confidence bounds were considered, they were equal.

The equal Weibull moduli for EW and LW at first point to similar underlying defect distributions for both tissues. In order to identify the underlying failure mechan-isms, representative failure faces of both EW and LW are given in Fig. 9. The middle lamella is considered the most likely point of failure when testing in T direction. This is confirmed by the micrographs of the failure faces for both EW and LW, as shown in Fig. 9 where the crack path follows the middle lamella.

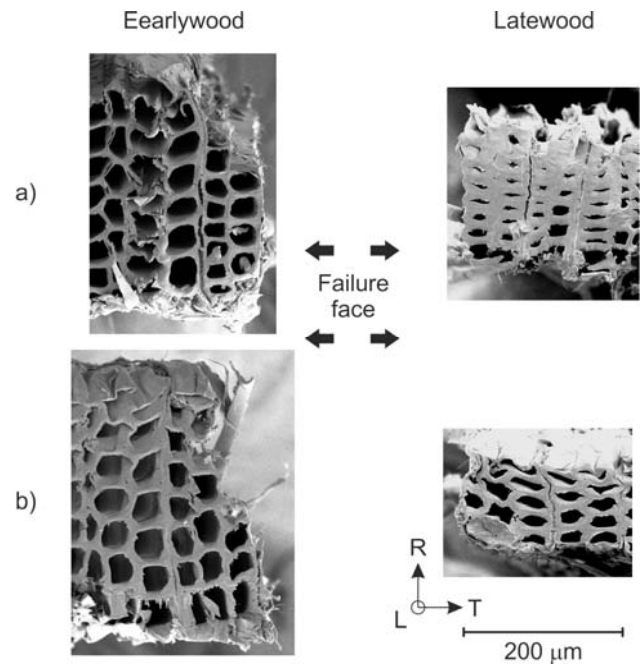

**Fig. 9.** Representative micrographs of the failure faces of EW (left) and LW (right) taken at a magnification of 600x.

These observations, however, are ex-situ and represent the failed material. Further in-situ investigations, which are undoubtedly challenging due to the brittle failure process, are needed to capture the failure process itself and shed light on the possible crack initiation function of the geometrical disorder which is introduced by pits and ray tissue.

This is valuable information with regard to, for example, the interpretation of bond line failures and crack propagation in wood. Furthermore, including imaging techniques allowed the associated failure mechanisms to be clearly identified.

**CONCLUSIONS**

1. The local mechanical parameters modulus of elasticity (MOE) and ultimate tensile stress (UTS) closely followed the density distribution within a growth ring. While the lowest values were found for the low-density earlywood (EW), the highest values were found for the high-density latewood (LW).

2. The influence of moisture content (MC) on the mechanical properties could only be assessed for EW. Although the results for the different tested MCs were not statistically significant due to their high variability, the mean values showed a clear decrease in mechanical properties with increasing MC for L and T direction.

3. The generic mechanical behavior of EW and LW in the T direction was evaluated on a separate set of samples, showing that although all strain-related quantities (strain at limit of elasticity, ultimate strain, and MOE) were significantly different for EW and LW, the stress-related quantities (stress at limit of elasticity and UTS) were not statistically different.

4. It was found that the non-linear stress/strain region was almost twice as large for EW as it was for LW.

5. The differences in the stress-related quantities could be associated with the different cell geometries; EW had a honeycomb-like cell arrangement and LW had a brick-like cell arrangement.
6. A Weibull analysis applied on the failure in T demonstrated that the underlying defect distributions in the T direction of EW and LW were similar. Images of the failure surfaces revealed the failure mechanisms: The failure occurred in the middle lamella for both, EW and LW.

**ACKNOWLEDGMENTS**

The funding by the Swiss National Science Foundation (Grant No. 125184) is gratefully acknowledged.

**REFERENCES**

Aicher, S., Höfflin, L., and Reinhardt, H.-W. (2007). "Runde Durchbrüche in Biegeträgern aus Brettschichtholz – Teil 2: Tragfähigkeit und Bemessung," *Bautechnik* 84(12), 867-880.

Astley, R. J., Stol, K. A., and Harrington, J. J. (1998). "Modelling the elastic properties of softwood." *Holz Roh. Werkst.*, 56(1), 43-50.

Burgert, I., Fruehmann, K., Keckes, J., Fratzl, P., and Stanzl-Tschegg, S. (2004). "Structure-function relationships of four compression wood types: Micromechanical properties at the tissue and fibre level," *Trees Struct. Fun.* 18(4), 480-485.

Boutelje, S. B. (1962). "The relationship of structure to transverse anisotropy," *Holzforschung* 16(2), 33-46.

Danielsson, H., and Gustafsson, P. (2010) "A probabilistic fracture mechanics method and strength analysis of glulam beams with holes," *Eur. J. Wood Wood Prod.* 69(3), 407-419.

Derome, D., Griffa, M., Koebel, M., and Carmeliet, J. (2011). "Hysteretic swelling of wood at cellular scale probed by phase-contrast X-ray tomography," *J. Struct. Biol.* 173(1), 180-190.

Derome, D., Rafsanjani, A., Hering, S., Dressler, M., Patera, A., Lanvermann, C., Sedighi Gilani, M., Wittel, F. K., Niemz, P., and Carmeliet, J. (2013). "The role of water in the behavior of wood," *J. Build. Phys.* 36(4), 398-421.

Dill-Langer, G., Lütze, S., and Aicher, S. (2002). "Microfracture in wood monitored by confocal laser scanning microscopy," *Wood Sci. Technol.* 36(6), 487-499.

DIN EN 61649 (2008). "Weibull analysis," DIN Deutsches Institut für Normung e. V.

Donaldson, L. (2008). "Microfibril angle: Measurement, variation and relationships," *Iawa J.* 29(4), 345-386.

Dvinskikh, S. V., Henriksson, M., Berglund, L. A., and Furó, I. (2011). "A multinuclear magnetic resonance imaging (MRI) study of wood with adsorbed water: Estimating bound water concentration and local wood density," *Holzforschung* 65(1), 103-107.

Eder, M., Jungnikl, K., and Burgert, I. (2009). "A close-up view of wood structure and properties across a growth ring of Norway spruce (*Picea abies* [L] Karst.)," *Trees-Struct. Funct.* 23(1), 79-84.

Ekevad, M., and Axelsson, A. (2012). "Variation of modulus of elasticity in the tangential direction with moisture content and temperature for Norway spruce (*Picea abies*)," *BioResources* 7(4), 4730-4743.

Evans, R., and Ilic, J. (2001). "Rapid prediction of wood stiffness from microfibril angle and density," *Forest Prod. J.* 51(3), 53-57.

Farruggia, F., and Perré, P. (2000). "Microscopic tensile tests in the transverse plane of earlywood and latewood parts of spruce," *Wood Sci. Technol.* 34(2), 65-82.

Fruehmann, K., Burgert, I., Stanzl-Tschegg, S. E., and Tschegg, E. K. (2003). "Mode I fracture behaviour on the growth ring scale and cellular level of spruce (*Picea abies* [L.] Karst.) and beech (*Fagus sylvatica* L.) loaded in the TR crack propagation system," *Holzforschung* 57(6), 653-660.

Gerhards, C. C. (1982). "Effect of moisture content and temperature on the mechanical properties of wood: An analysis of immediate effects," *Wood Fiber Sci.* 14(1), 4-36.

Gustafsson, P. (2003). "Fracture perpendicular to grain – structural applications," in: *Timber Engineering*, John Wiley & Sons Ltd., Chichester, 103-130.

Hass, P., Wittel, F. K., Miller, M., and Niemz, P. (2012). "Generic failure mechanisms in adhesive bonds," *Holzforschung* 67(2), 207-215 .

Jernkvist, L. O. and Thuvander, F. (2001). "Experimental determination of stiffness variation across growth rings in *Picea abies*," *Holzforschung* 55(3), 309-317.

Kahle, E., and Woodhouse, J. (1994). "The influence of cell geometry on the elasticity of softwoods," *J. Mater. Sci.* 29(5), 1250-1259.

Keunecke, D., Hering, S., and Niemz, P. (2008). "Three-dimensional elastic behaviour of common yew and Norway spruce," *Wood Sci. Technol.* 42(8), 633-647.

Keunecke, D., Novosseletz, K., Lanvermann, C., Mannes, D., and Niemz, P. (2012). "Combination of X-ray and digital image correlation for the analysis of moisture-induced strain in wood: Opportunities and challenges," *Eur J. Wood Wood Prod.* 70(4), 407-413.

Kifetew, G. (1999). "The influence of the geometrical distribution of cell-wall tissues on the transverse anisotropic dimensional changes of softwood." *Holzforschung*, 53(4), 347-349.

Lanvermann, C., Sanabria, S. S., Mannes, D., and Niemz, P. (2013a). "Combination of neutron imaging and digital image correlation to determine intra-ring moisture variation in norway spruce," *Holzforschung Online First*

Lanvermann, C., Schmitt, U., Evans, R., Hering, S., and Niemz, P. (2013b). "Distribution of structure and lignin within growth rings of Norway spruce," *Wood Sci. Technol.* 47(3), 627-641.

Modén, C. S., and Berglund, L. A. (2008a). "Elastic deformation mechanisms of softwoods in radial tension - Cell wall bending or stretching?" *Holzforschung* 62(5), 562-568.

Modén, C. S., and Berglund, L. A. (2008b). "A two-phase annual ring model of transverse anisotropy in softwoods," *Compos. Sci. Technol*, 68(14), 3020-3026.

Moon, R. J., Wells, J., Kretschmann, D. E., Evans, J., Wiedenhoeft, A. C., and Frihart, C. R. (2010). "Influence of chemical treatments on moisture-induced dimensional change and elastic modulus of earlywood and latewood," *Holzforschung* 64(6), 771-779.

Nanjangud, S. C., Brezny, R., and Green, D. J. (1995). "Strength and Young's modulus behavior of a partially sintered porous alumina," *J. Am. Ceram. Soc.* 75(1), 266-268.

Neuhaus, H. (1983). "Über das elastische Verhalten von Fichtenholz in Abhängigkeit von der Holzfeuchtigkeit," *Holz Roh. Werkst.* 41(1), 21-25.

Naylor, A., Hackney, P., Perera, N., and Clahr, E. (2012). "A predictive model for the cutting force in wood machining developed using mechanical properties," *BioResources* 7(3), 2883-2894.

Nyahumwa, C. (2005). "Multiple defect distributions on Weibull statistical analysis of fatigue life of cast aluminium alloys," *African Journal of Science and Technology* 6(2), 43-54.

Persson, K. (2000). *Micromechanical Modelling of Wood and Fibre Properties*, PhD Thesis, Lund University.

Rafsanjani, A., Derome, D., Wittel, F. K., and Carmeliet, J. (2012). "Computational up-scaling of anisotropic swelling and mechanical behavior of hierarchical cellular materials," *Compos. Sci. Technol*. 72(6), 744-751.

Rosner, S. (2012). "Waveform features of acoustic emission provide information about reversible and irreversible processes during spruce sapwood drying," *BioResources* 7(1), 1253-1263.

Roszyk, E., Moliński, W., and Fabisiak, E. (2013). "Radial variation of mechanical properties of pine wood (*Pinus sylvestris* L.) determined upon tensile stress," *Wood Research* 58(3), 329-342.

Sinn, G., Reiterer, A., Stanzl-Tschegg, S. E., and Tschegg, E. K. (2001). "Determination of strains of thin wood samples using videoextensometry," *Holz Roh. Werkst.* 59(3), 177-182.

Skaar, S. (1988), *Wood-Water Relations*, Springer-Verlag, Heidelberg.

Wagenführ, R. (2000). *Holzatlas*, Fachbuchverlag, Leipzig.

Watanabe, U. (1998). "Shrinking and elastic properties of coniferous wood in relation to cellular structure," *Wood Research* 85, 1-47.

Weibull, W. (1951). "A statistical distribution function of wide applicability," *J. Appl. Mech*. 18(3), 293-297.

Wittel, F. K., Dill-Langer, G., and Kröplin, B. (2005). "Modeling of damage evolution in soft-wood perpendicular to grain by means of a discrete element approach," *Comp. Mat. Sci*. 32(3-4), 594-603.